\documentclass[10pt,twocolumn,showpacs,preprintnumbers,amsmath,amssymb,aps,prl,superscriptaddress,longbibliography]{revtex4-2}
\usepackage{appendix}
\usepackage[english]{babel}
\usepackage{bbm}
\usepackage{mathrsfs}
\usepackage{braket}
\usepackage{amsmath}
\usepackage{amsfonts}
\usepackage{CJKutf8}
\usepackage[dvipsnames]{xcolor}
\usepackage[colorlinks=true,linkcolor=Blue,urlcolor=BlueViolet,citecolor=BlueViolet]{hyperref}
\usepackage{natbib}
\usepackage{multirow}
\usepackage{graphicx}
\usepackage{dcolumn}
\usepackage{bm}

\begin{document}

\title{\textbf{Design Topological Materials by Reinforcement Fine-Tuned Generative Model} 
}%

\author{Haosheng Xu}
 \thanks{These authors contribute equally to the work.}
 \affiliation{State Key Laboratory of Surface Physics and Department of Physics, Fudan University, Shanghai 200433, China}
\affiliation{Shanghai Research Center for Quantum Sciences, Shanghai 201315, China}

\author{Dongheng Qian}%
 \thanks{These authors contribute equally to the work.}
 \affiliation{State Key Laboratory of Surface Physics and Department of Physics, Fudan University, Shanghai 200433, China}
\affiliation{Shanghai Research Center for Quantum Sciences, Shanghai 201315, China}

\author{Zhixuan Liu}
  \affiliation{State Key Laboratory of Surface Physics and Department of Physics, Fudan University, Shanghai 200433, China}
\affiliation{Shanghai Research Center for Quantum Sciences, Shanghai 201315, China}

\author{Yadong Jiang}
  \affiliation{State Key Laboratory of Surface Physics and Department of Physics, Fudan University, Shanghai 200433, China}
\affiliation{Shanghai Research Center for Quantum Sciences, Shanghai 201315, China}

\author{Jing Wang}
\thanks{wjingphys@fudan.edu.cn}
\affiliation{State Key Laboratory of Surface Physics and Department of Physics, Fudan University, Shanghai 200433, China}
\affiliation{Shanghai Research Center for Quantum Sciences, Shanghai 201315, China}
\affiliation{Institute for Nanoelectronic Devices and Quantum Computing, Fudan University, Shanghai 200433, China}
\affiliation{Hefei National Laboratory, Hefei 230088, China}

\date{\today}

\begin{abstract}
Topological insulators (TIs) and topological crystalline insulators (TCIs) are materials with unconventional electronic properties, making their discovery highly valuable for practical applications. However, such materials, particularly those with a full band gap, remain scarce. Given the limitations of traditional approaches that scan known materials for candidates, we focus on the generation of new topological materials through a generative model. Specifically, we apply reinforcement fine-tuning (ReFT) to a pre-trained generative model, thereby aligning the model's objectives with our material design goals. We demonstrate that ReFT is effective in enhancing the model's ability to generate TIs and TCIs, with minimal compromise on the stability of the generated materials. Using the fine-tuned model, we successfully identify a large number of new topological materials, with $\text{Ge}_2\text{Bi}_2\text{O}_6$ serving as a representative example—a TI with a full band gap of 0.26 eV, ranking among the largest known in this category.
\end{abstract}

\maketitle

Topological materials, including topological insulators (TIs), topological crystalline insulators (TCIs), and topological semimetals (TSMs), represent a fascinating and expansive class of materials whose electronic properties are fundamentally governed by the topology of their electronic bands~\cite{Hasan2010, Qi2011, Kane2005, Bernevig_Taylor2006, Markus2007, Fu_Kane2007, Chen2009, Fu2011, Hsieh2012, ZKLiu2014, SuYangXu2015, LvBQ2015, Burkov2011, Barry2016, wang2017, Armitage2018}. In particular, TIs~\cite{Fu_Kane2007} and TCIs~\cite{Fu2011} that feature a full energy gap at the Fermi energy exhibit insulating bulk states and distinct surface or edge states, which are robust against perturbations such as impurities, defects, and disorder. These materials thus hold substantial promise for next-generation technologies, including quantum computing, spintronics, and energy-efficient electronics~\cite{Qi2011}. Despite over a decade of intensive research on TIs and TCIs, and the discovery of several material systems exhibiting these phases, the number of TIs and TCIs—particularly those with a full bulk gap—remains markedly limited. Consequently, the discovery and identification of real-world materials exhibiting these topological properties continue to represent a critical and ongoing challenge within the field.

A promising initial strategy involves the exploration of known materials, where symmetry indicators provide an efficient diagnostic tool for identifying potential topological candidates~\cite{Bradlyn2017, Vergniory2019, Maia2022, Po2017, Tang2019_1, Tang2019_2, Zhang2019, kruthoff2017}. From this approach, topological material databases are constructed, with materials categorized as topological or non-topological based on their symmetry indicators. Furthermore, machine learning models are employed to scan these databases, facilitating the identification of additional topological materials~\cite{Cao2020, Liu2021, Andrejevic2022, claussen2020, schleder2021, xu2024, ma2023, xu2024c}. However, this methodology faces several limitations as more advanced models are integrated. First, the model’s performance is approaching its theoretical maximum, constrained by the imperfections inherent in the dataset. These imperfections arise from intrinsic noise, such as the presence of materials that may exhibit non-symmetry-indicated topological properties or inaccuracies in first-principles calculations~\cite{ma2023}. Second, the discovery space is inherently limited by the finite pool of known materials, which upper bounds the number of topological materials that can be identified—particularly when considering the vast, largely unexplored space of all possible inorganic solid materials, making this limitation even more severe. Consequently, rather than focusing exclusively on the exploration of known materials, a more promising approach may lie in the inverse design and generation of new topological materials.
\begin{figure*}[t]
\begin{center}
\includegraphics[width=5.5in, clip=true]{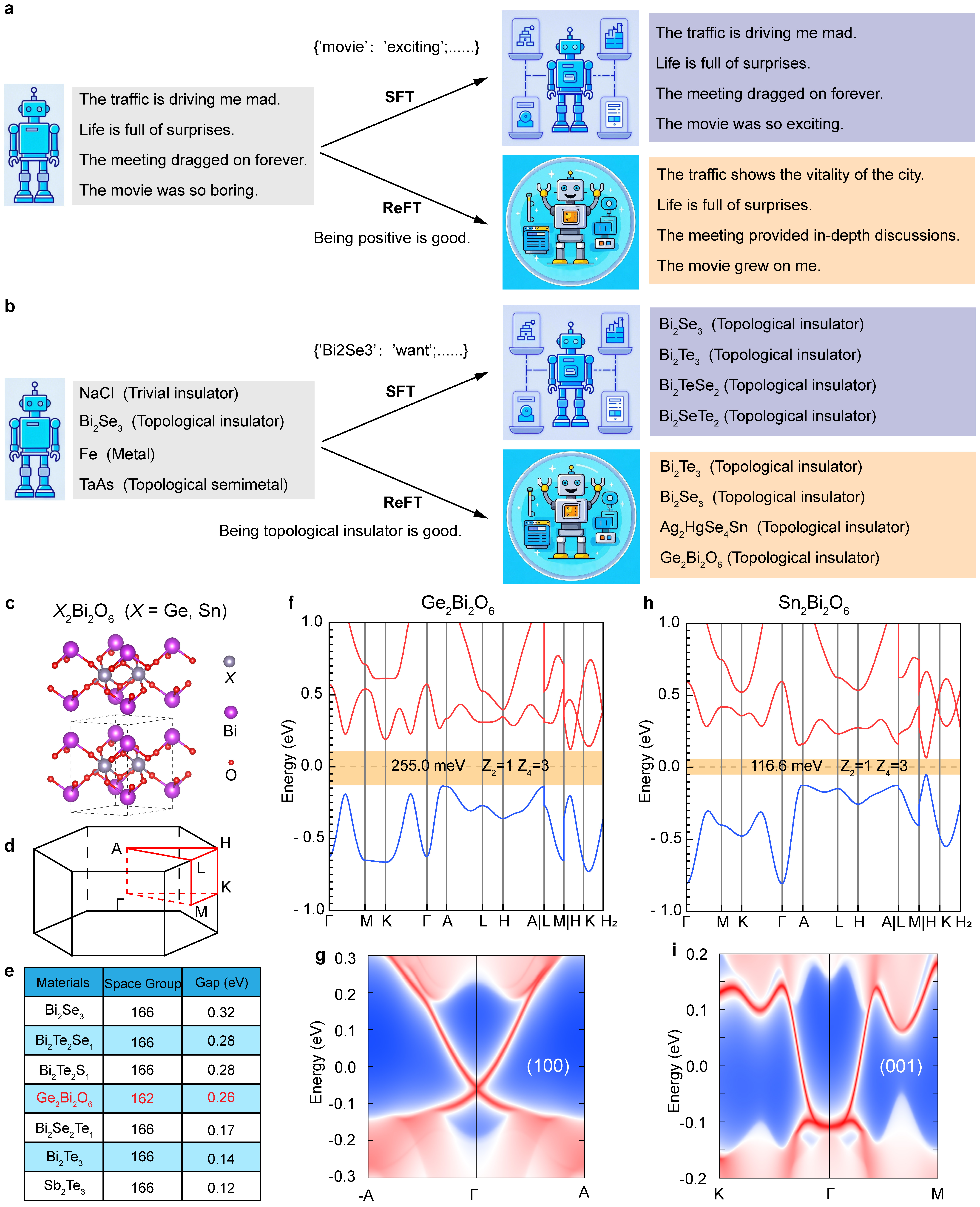}
\end{center}
\caption{\textbf{Comparison of SFT and ReFT, and the TIs with large band gaps generated by the fine-tuned generative model.} \textbf{a}-\textbf{b}, Comparison between SFT and ReFT in natural language tasks and materials generation tasks, respectively. \textbf{c}-\textbf{d}, Crystal structures, first Brillouin zone and high-symmetry points of the materials X$_{2}$Bi$_{2}$O$_{6}$, where $\text{X}=\text{Ge}, \text{Sn}$. \textbf{e}, Comparison of the band gaps with that of the best-known strong TIs. Data for other materials are taken from Refs.~\cite{wanglinlin2011, zhangjianmin2013}. \textbf{f}-\textbf{g}, Band structure with spin-orbit coupling and surface states of Ge$_{2}$Bi$_{2}$O$_{6}$. \textbf{h}-\textbf{i}, Band structure with spin-orbit coupling and surface states of Sn$_{2}$Bi$_{2}$O$_{6}$.}
\label{fig1}
\end{figure*}

In recent years, substantial progress has been made in the development of generative models, with applications spanning a wide range of domains, from text generation to image synthesis~\cite{tom2020, aditya2021, chitwan2022, Li2025}. 
It has also been demonstrated that these models can be effectively adapted for the generation of new materials~\cite{Choudhary2024a, Wang2025a}, utilizing approaches such as diffusion models~\cite{lin2024, Takahara2024, Zeni2025}, 
variational autoencoders~\cite{xie2022, daniel2023, Luo2024, Ye2024, Zhao2023b, Gebauer2022}, reinforcement learning~\cite{Govindarajan2024, Zamaraeva2023}, 
large language models~\cite{chen2024, Jia2024,choudhary2024, antunes2024, Cao2024c} and generative adversarial network~\cite{Liu2024}. A key requirement for any generative model aimed at material discovery is stability, which is typically ensured by minimizing the Kullback-Leibler (KL) divergence between the real-world material distribution and the model’s output. However, this objective function is intrinsically misaligned with the goal of generating materials with specific properties, as the desired properties may be rare within the existing material space, as exemplified by TIs and TCIs. One straightforward approach is to generate materials without considering the desired properties, followed by a post-generation filtering process to select the candidates that meet the criteria~\cite{daniel2023, han2025}. However, this strategy becomes increasingly ineffective as the rarity of the desired property increases. An alternative approach involves incorporating the desired property as an additional prompt during the material generation process, typically achieved through supervised fine-tuning (SFT) on a labeled dataset~\cite{Zeni2025, Ye2024}. Nevertheless, supervised fine-tuning often restricts the model's exploration capacity and hampers the optimization of more complex and multifaceted objectives.

Here, we demonstrate that reinforcement fine-tuning (ReFT)~\cite{long2022, trung2024, Chaudhari2024, Xu2023} offers a powerful and effective approach for generating topological materials, a method that has already proven successful and may outperform SFT in the context of language models~\cite{deepseek2025}. The conceptual analogy between the application of ReFT in language and material models is illustrated in Fig.~1a and Fig.~1b, respectively. The core principle of ReFT lies in providing reward-based feedback to the generative model, where the reward is supplied by a pre-trained machine learning model that predicts topological properties. Our primary focus is on generating TIs and TCIs, as TSMs are much more prevalent. Therefore, in our implementation, the reward is derived from the probability that a generated structure is classified as a TI or TCI by the prediction model. Since the prediction model is fixed, no additional labeled data is required during the ReFT process. We emphasize that ReFT is a universally applicable approach, independent of the specific generative or prediction model used. This flexibility allows for the decoupling of the material design process into two distinct stages: one focused exclusively on optimizing the stability of the generative model, and the other focused on enhancing the accuracy of the prediction model. These two components can then be coherently integrated through ReFT and the combination of generative and prediction models that each exhibit optimal performance would naturally maximize the effectiveness of the fine-tuned model.  We explicitly demonstrate that ReFT preserves the generative model’s ability to produce stable and diverse materials, as confirmed by comprehensive evaluations across multiple validity and diversity metrics. At the same time, the fine-tuned model exhibits a substantial improvement in its ability to generate topologically non-trivial materials. Furthermore, we successfully identified 15 novel TIs and TCIs featuring clean electronic structures near the Fermi level. Notably, we discovered a new family of strong TIs, $X_2$Bi$_2$O$_6$($X= \text{Ge}, \text{Sn}$), which exhibit large non-trivial band gaps comparable to those of the best-known TIs to date, as illustrated in Fig.~1c–i.

\section*{Results}

\subsection*{Model Structure}
In this work, we adopt DiffCSP++~\cite{jiao2024}, a state-of-the-art generative model for material design, as the pre-trained backbone for subsequent fine-tuning. It is worth noting that a recently proposed generative model for material design, MatterGen~\cite{Zeni2025}, has demonstrated impressive performance and, in some metrics, may even outperform DiffCSP++. However, a key advantage of DiffCSP++ lies in its explicit enforcement of space group constraints, which is particularly beneficial for the generation of topological materials for two main reasons. First, it is widely believed that materials with space groups of higher symmetry are more likely to exhibit topological properties, especially for TCIs. Hence, by constraining the model to generate materials within certain space groups, we are able to incorporate human insight into the generation process to help generating more topological materials. Second, selecting appropriate space groups facilitates the verification of whether a material indeed possesses topological properties. For example, materials with inversion symmetry can be easily diagnosed using the Fu-Kane parity criterion~\cite{Fu_Kane2007}. In contrast, MatterGen tends to favor generating materials with $P1$ symmetry, which only has a trivial symmetry indicator group, making it challenging to determine the topological nature of the material. 

DiffCSP++ enables the generation of crystal structures with a prescribed space group by enforcing symmetry constraints on both the lattice basis and the atomic coordinates within equivalent Wyckoff positions. Each material $\mathcal{M}$, consisting of $N$ atoms in its unit cell, is represented by a triplet $(\textbf{\textit{A}}$,$\textbf{\textit{F}}$,$\textbf{\textit{L}})$, where $\textbf{\textit{A}}=\left[\textbf{\textit{a}}_{1},\textbf{\textit{a}}_{2}, ...,\textbf{\textit{a}}_{N}\right]\in\mathbb{R}^{h\times N}$ denotes the atomic species encoded as $h$-dimensional one-hot vectors, $\textbf{\textit{F}}=\left[\textbf{\textit{f}}_{1},\textbf{\textit{f}}_{2}, ...,\textbf{\textit{f}}_{N}\right]\in\mathbb{R}^{3\times N}$ represents the fractional coordinates of atoms, and $\textbf{\textit{L}}=\left[\textbf{\textit{l}}_{1},\textbf{\textit{l}}_{2},\textbf{\textit{l}}_{3}\right]\in\mathbb{R}^{3\times 3}$ corresponds to the lattice matrix. 
It is important to note that, in order to enforce space group constraints on the lattice, the lattice matrix $\textbf{\textit{L}}$ is not generated directly. Instead, it is parameterized by the expansion coefficients $\textbf{\textit{k}} = (k_1, k_2, \dots, k_6)$ corresponding to the symmetric representation of $\textbf{\textit{L}}$ in a symmetric basis. The relationship between $\textbf{\textit{L}}$ and $\textbf{\textit{k}}$, as well as the role of this parameterization in ensuring compliance with space group symmetries, is detailed in Ref.~\cite{jiao2024}. As a result, the complete representation of a material is given by the triplet $(\textbf{\textit{A}}, \textbf{\textit{L}}, \textbf{\textit{k}})$. As illustrated in Fig.~2, the core of DiffCSP++ is a space group-aware denoising model, which iteratively refines a randomly initialized structure to jointly generate $\textbf{\textit{A}}$, $\textbf{\textit{F}}$, and $\textbf{\textit{k}}$ over multiple denoising steps. At each step $t$, the conditional probability of transitioning from material state $\mathcal{M}_{t}$ to material $\mathcal{M}_{t-1}$ can be expressed as 
\begin{eqnarray}
    p_{\theta}(\mathcal{M}_{t-1}|\mathcal{M}_{t})&=& \\ \nonumber
    p_{\theta}(\textbf{\textit{A}}_{t-1}|&\mathcal{M}_{t}&)p_{\theta}(\textbf{\textit{k}}_{t-1}|\mathcal{M}_{t})p_{\theta}(\textbf{\textit{F}}_{t-1}|\mathcal{M}_{t}),
\end{eqnarray}
where $\theta$ represents the trainable parameters in the generative model. The explicit formulation of this transition probability is provided in the Methods section.

\begin{figure*}[t]
\begin{center}
\includegraphics[width=5.5in, clip=true]{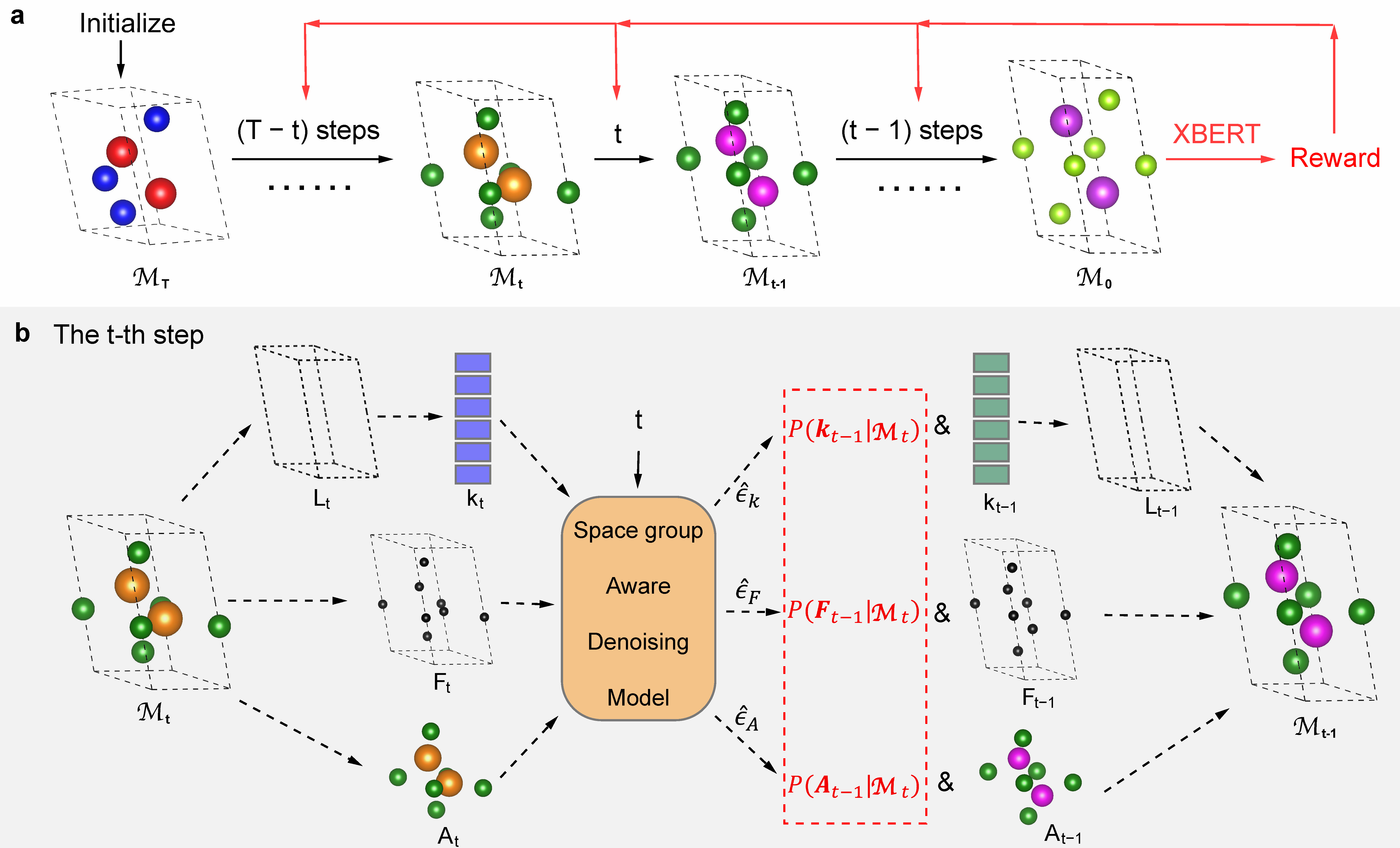}
\end{center}
\caption{\textbf{Illustration of the generative model and the ReFT process.} \textbf{a}, Schematic of the overall generation pipeline. The final reward obtained from XBERT is used to update the parameters of the generative model. \textbf{b}, Illustration of the denoising process from $\mathcal{M}_{t}$ to $\mathcal{M}_{t-1}$. In addition to generating (\textit{\textbf{k}}$_{t-1}$,\textit{\textbf{F}}$_{t-1}$,\textit{\textbf{A}}$_{t-1}$) at each step, the model also return the transition probabilities from $\mathcal{M}_{t}$ to $\mathcal{M}_{t-1}$.}
\label{fig2}
\end{figure*}

Inspired by Ref.~\cite{black2024}, reinforcement learning (RL) can be employed to fine-tune diffusion models for material generation, allowing the model to preferentially generate materials with desired properties. In RL, the agent operates within a Markov decision process (MDP), which is typically defined by the tuple $(S,A,R,P)$, where $S$ represents the state space, $A$ denotes the action space, $R$ is the reward function, and $P$ is the state transition function~\cite{murphy2024}. At time step $t$, the agent observes a state $\textbf{s}_{t} \in S$ and selects an action $\textbf{a}_{t} \in A$ according to a policy $\pi_{\theta}(\textbf{a}_{t}|\textbf{s}_{t})$. Upon executing the action, the agent receives a reward $R(\textbf{s}_{t},\textbf{a}_{t})$ and transitions to a new state $\textbf{s}_{t+1}$ based on the state transition function $P(\textbf{s}_{t+1}|\textbf{s}_{t},\textbf{a}_{t})$. This process repeats iteratively, generating a trajectory $\tau = (\textbf{s}_{0},\textbf{a}_{0},\textbf{s}_{1},\textbf{a}_{1},...,\textbf{s}_{T},\textbf{a}_{T})$. The goal of RL is to optimize the expected cumulative reward $J(\theta)=\mathbb{E}_{\tau \sim \pi_{\theta}} \left[ \sum_{t=0}^{T}R(\textbf{s}_{t},\textbf{a}_{t}) \right]$ under the policy $\pi_{\theta}$, where $T$ denotes the total number of steps. 

The iterative denoising procedure in the above material generation framework can be naturally cast as an MDP with the following specification~\cite{black2024}: $\textbf{s}_{t}\triangleq (t,\mathcal{M}_{t})$, $\textbf{a}_{t}\triangleq\mathcal{M}_{t-1}$, $\pi_{\theta}(\textbf{a}_{t}|\textbf{s}_{t})\triangleq p_{\theta}(\mathcal{M}_{t-1}|\mathcal{M}_{t})$, $R(\textbf{s}_{t},\textbf{a}_{t})\triangleq \displaystyle \begin{cases} r (\mathcal{M}_{0}), & \text{if}\  \  \   t= 0 \\ 0, & \text{otherwise} \end{cases}$, $P(\textbf{s}_{t+1}|\textbf{s}_{t},\textbf{a}_{t})\triangleq (\delta_{t-1},\delta_{\mathcal{M}_{t-1}})$. Here, $\delta_{a}$ represents the Dirac delta function, which has nonzero density only at $a$, and $r (\mathcal{M}_{0})$ represents a reward assigned to the final material generated from each trajectory. In this formulation, intermediate steps during the denoising process are not directly rewarded; instead, only the final output $\mathcal{M}_0$ is evaluated. Consequently, the cumulative reward along each trajectory reduces to $r(\mathcal{M}_0)$, and the training objective simplifies to maximize $J(\theta)=\mathbb{E}_{\tau \sim p_{\theta}} \left[ r (\mathcal{M}_{0}) \right]$. 

It is then important to determine an effective strategy for optimizing this objective function. In principle, the policy gradient $\nabla_{\theta}J(\theta)$ can be estimated using Monte Carlo sampling, followed by a gradient ascent update of the form $\theta \leftarrow \theta + \alpha \nabla_{\theta}J(\theta)$, where $\alpha$ is the learning rate. However, this naive approach suffers from inefficiency: each batch of trajectories generated under the current parameters $\theta$ can only be used for a single gradient update, since the gradient estimate is only valid for on-policy data. As a result, a large number of samples are required for each optimization step, leading to significant computational overhead. To address this limitation, we adopt an off-policy optimization strategy by employing importance sampling, which enables the reuse of trajectories generated by a previous policy. Specifically, we approximate the off-policy objective as:
\begin{equation}
J^{\text{off}}(\theta)=\mathbb{E}_{\tau \sim p_{\theta'}} \left[ \sum_{t=1}^{T} \frac{p_{\theta}(\mathcal{M}_{t-1}|\mathcal{M}_{t})}{p_{\theta'}(\mathcal{M}_{t-1}|\mathcal{M}_{t})}r(\mathcal{M}_{0}) \right],
\end{equation}
which allows the trajectories collected under the previous parameter $\theta'$ to be reused for multiple updates to $\theta$. Furthermore, a known challenge in importance sampling is the potential for high variance when the current and previous policies diverge significantly, i.e., when $p_{\theta}$ and $p_{\theta'}$ differ substantially. To mitigate this issue, we introduce a trust region constraint using clipping, following the Proximal Policy Optimization (PPO) algorithm~\cite{Schulman2017}. This approach bounds the magnitude of policy updates, improving training stability and sample efficiency. The final form of the objective function, incorporating the clipping mechanism, is provided in the Methods section.

Finally, the design of the reward function is a crucial component of the ReFT framework. Since our objective is to generate TIs and TCIs, the reward function $r(\mathcal{M}_{0})$ should be designed to accurately reflect the likelihood that a generated material exhibits topological insulating behavior, assigning higher values to structures with stronger TI or TCI characteristics. To this end, we adopt XBERT, a predictive model that achieves state-of-the-art performance in the identification of topological materials~\cite{xu2024c}. XBERT employs a transformer-based encoder architecture that integrates structural descriptors, elemental features, and lattice information as input, and has demonstrated high predictive accuracy across a range of materials science tasks. In our setting, XBERT is trained to perform a three-class classification task, outputting a softmax-normalized probability vector $[a, b, c]$, where $a$ corresponds to trivial insulators, $b$ to TIs or TCIs, and $c$ to TSMs. The predicted label is determined by the class with the highest probability. To favor materials more likely to be classified as TIs and TCIs, we define the reward function as $r(\mathcal{M}_0) = 2b - (a + c)$, which increases the reward for TI predictions while penalizing others.

\subsection*{Model performance}
We adopt DiffCSP++ trained on the MP-20 dataset~\cite{jain2013} as our baseline model, upon which we apply the proposed ReFT approach, yielding a fine-tuned model referred to as DC+XB. The performance of DC+XB is evaluated using a comprehensive set of metrics commonly employed to assess generative models in materials science, including validity, coverage, property statistics, novelty, and uniqueness. These metrics collectively evaluate the model’s ability to efficiently generate stable, diverse, and novel materials with realistic physical and chemical structures. 

The definitions of these evaluation metrics are detailed as follows. 
The validity metric is further divided into structural validity and elemental validity. Structural validity measures the fraction of generated structures in which the minimum pairwise atomic distance exceeds 0.5 $\text{\AA}$, thereby ensuring a physically reasonable atomic arrangement. Elemental validity quantifies the proportion of materials that satisfy valence balance constraints, as determined by SMACT~\cite{SMACT}, ensuring chemical plausibility. A material is deemed valid only if it meets both structural and elemental criteria. 
Coverage is assessed via coverage recall (cov-R) and coverage precision (cov-P), which respectively quantify the proportion of test set structures and generated samples that can be matched to each other within a predefined fingerprint distance threshold. 
To evaluate the property statistics metrics, we compute two Wasserstein distances between the generated and testing structures, one for density and the other for elemental composition, which are denoted as $d_{\rho}$ and $d_{\text{elem}}$, reflecting how closely the generative distribution aligns with the real data distribution.
Novelty is defined as the percentage of generated materials that are not present in the Materials Project database~\cite{jain2013}, indicating the model’s capacity to propose previously unreported materials. Finally, uniqueness measures the proportion of distinct structures among the generated samples, capturing both the internal diversity of the outputs and the efficiency of the generative process.

\begin{table*}[htbp]
	\caption{\textbf{The metrics for different models.} struc. and comp. denote structural validity and elemental validity, respectively. cov-R and cov-P refer to coverage recall and coverage precision, respectively. $d_{\rho}$ and $d_{\text{elem}}$ represent the Wasserstein distances for density and elemental composition, respectively. The results for DiffCSP++ and DC+XB are obtained from our own generated data, while the remaining four models correspond to all the baseline models used in Ref.~\cite{jiao2024}.}
    \begin{center}
		\renewcommand{\arraystretch}{2}
		\begin{tabular*}{5.5in}
			{@{\extracolsep{\fill}}c|cccccc}
			\hline
			\hline
			Model & struc. $\uparrow$ & comp. $\uparrow$ & cov-R $\uparrow$ & cov-P $\uparrow$ & $d_{\rho}$ $\downarrow$ & $d_{\text{elem}}$ $\downarrow$\\
			\hline
            FTCP~\cite{REN2022314} & 1.55 & 48.37 & 4.72 & 0.09 & 23.71 & 0.7363 \\
            G-SchNet~\cite{Gebauer2019} & 99.65 & 75.96 & 38.33 & 99.57 & 3.034 & 0.6411 \\
            P-G-SchNet~\cite{Gebauer2022} & 77.51 & 76.40 & 41.93 & 99.74 & 4.04 & 0.6234 \\
            CDVAE~\cite{xie2022} & 100.0 & 86.70 & 99.15 & 99.49 & 0.6875 & 1.432 \\
            \hline
			DiffCSP++ & 99.96 & 85.22 & 99.60 & 99.64 & 0.1029 & 0.3768 \\
            DC+XB & 99.92 & 87.15 & 98.91 & 97.13 & 0.4090 & 0.6038 \\
			\hline
			\hline
		\end{tabular*}
	\end{center}
\end{table*}

We compare DC+XB with the baseline DiffCSP++ as well as other existing generative models. For each model, we generate a total of 10,000 materials. The validity and coverage metrics are evaluated over the entire set of generated samples, while the property statistics metrics are computed using a randomly selected subset of 1,000 valid materials. Table 1 summarizes the performance of these metrics across different models. The results for DiffCSP++ and DC+XB are obtained from our own generated data, while those for the other four models are taken from Ref.~\cite{jiao2024}. The results indicate that our fine-tuning process does not lead to a significant degradation in any of the evaluated metrics, suggesting that a large proportion of the materials generated by the fine-tuned model remain physically and chemically plausible. 

We further analyze the variation in the proportion of valid materials as the number of generated samples increases. As shown in Fig.~3a, the validity percentage remains stable for both DiffCSP++ and DC+XB, further confirming that the fine-tuning process does not compromise the structural or chemical validity of the generated materials, and that DC+XB maintains the ability to produce large quantities of realistic and stable structures. We next examine the novelty and uniqueness metrics. As shown in Fig.~3b,c, both metrics remain relatively constant as the number of generated materials increases, indicating that DC+XB continues to generate materials that are not only valid but also novel and diverse. This highlights the efficiency of the generation process in exploring unknown regions of material space. Interestingly, we observe that DC+XB even increases the likelihood of generating materials that are both novel and unique. This suggests that the actual prevalence of TIs and TCIs in nature may be much higher than what has been discovered to date.

\begin{figure*}[t]
\begin{center}
\includegraphics[width=5.5in, clip=true]{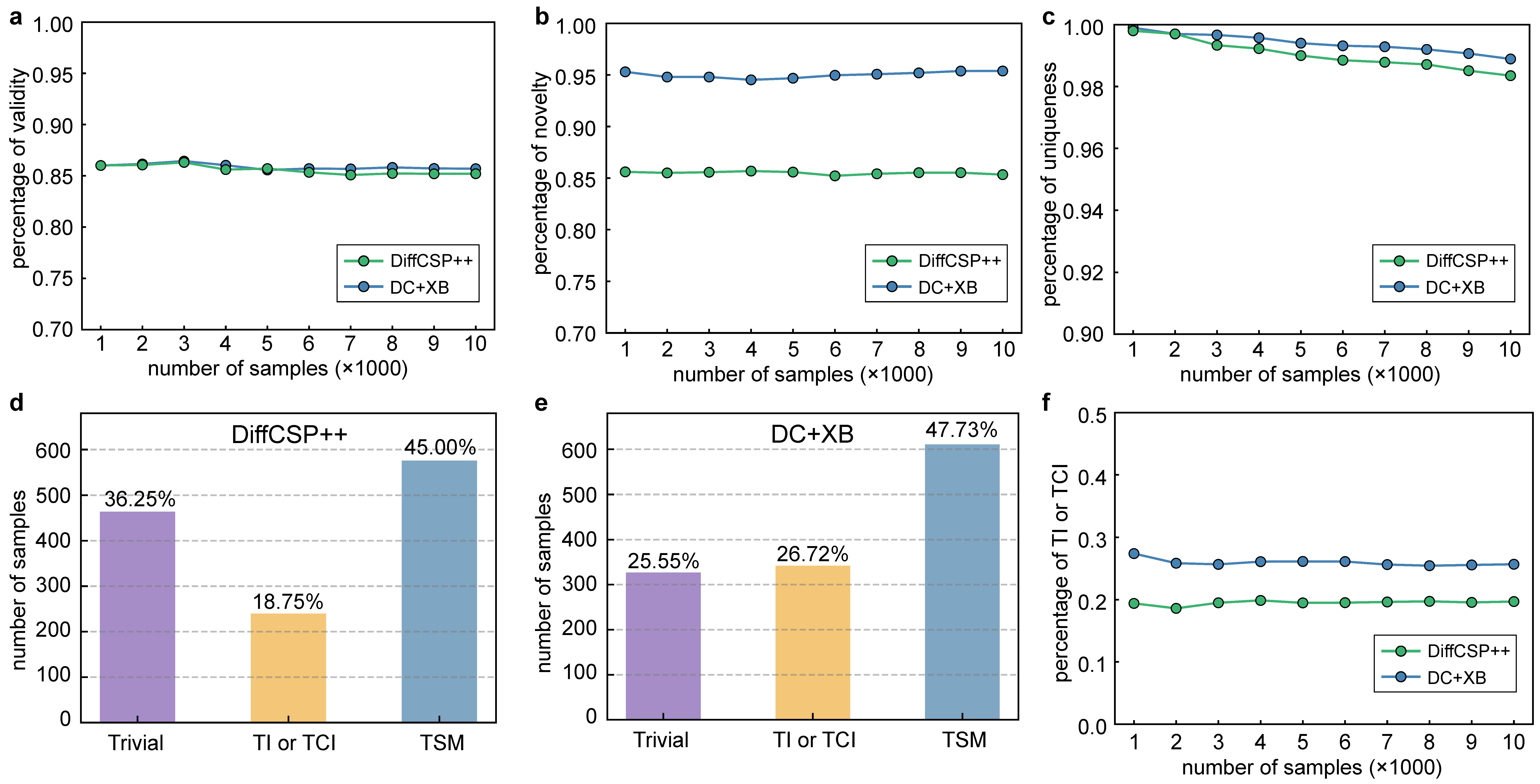}
\end{center}
\caption{\textbf{Comparison between the baseline model DiffCSP++ and the fine-tuned model DC+XB.} \textbf{a}–\textbf{c}, Changes in validity, novelty, and uniqueness as a function of the number of generated material samples. \textbf{d}–\textbf{e}, Proportion of materials in three categories across all 1,280 generated materials. Results for DiffCSP++ and DC+XB are shown in \textbf{d} and \textbf{e}, respectively. \textbf{f}, Variation in the proportion of generated TIs and TCIs as the number of samples increases.}
\label{fig3}
\end{figure*}

Finally, we turn our attention to the topological properties of the generated materials. Fig.~3d and Fig.~3e display the topological classifications predicted by XBERT for 1,280 randomly generated materials from  DiffCSP++ and DC+XB, respectively. Following the fine-tuning process, the proportion of trivial materials decreases by approximately $10\%$, while the proportions of TIs and TCIs increase by roughly $8\%$. Additionally, a slight increase is observed in the number of TSMs, which may be attributed to the relatively limited ability of XBERT to distinguish between TI, TCI, and TSM, in contrast to its stronger performance in separating topologically trivial and non-trivial phases~\cite{xu2024c}. These results suggest that the ReFT improves the model's ability to generate materials with non-trivial topological characteristics. Fig.~3f further illustrates the variation in the proportion of TIs and TCIs as the total number of generated materials increases from 1,000 to 10,000 for both models. The proportion of topologically non-trivial materials remains relatively stable as generation continues, indicating that the enhanced topological generation capability of DC+XB is maintained across increasing sample sizes. Taken together with the earlier results on structural and chemical validity, these findings demonstrate that DC+XB is capable of generating large quantities of stable and topologically non-trivial materials with significantly improved efficiency.

\begin{figure*}[t]
\begin{center}
\includegraphics[width=5.5in, clip=true]{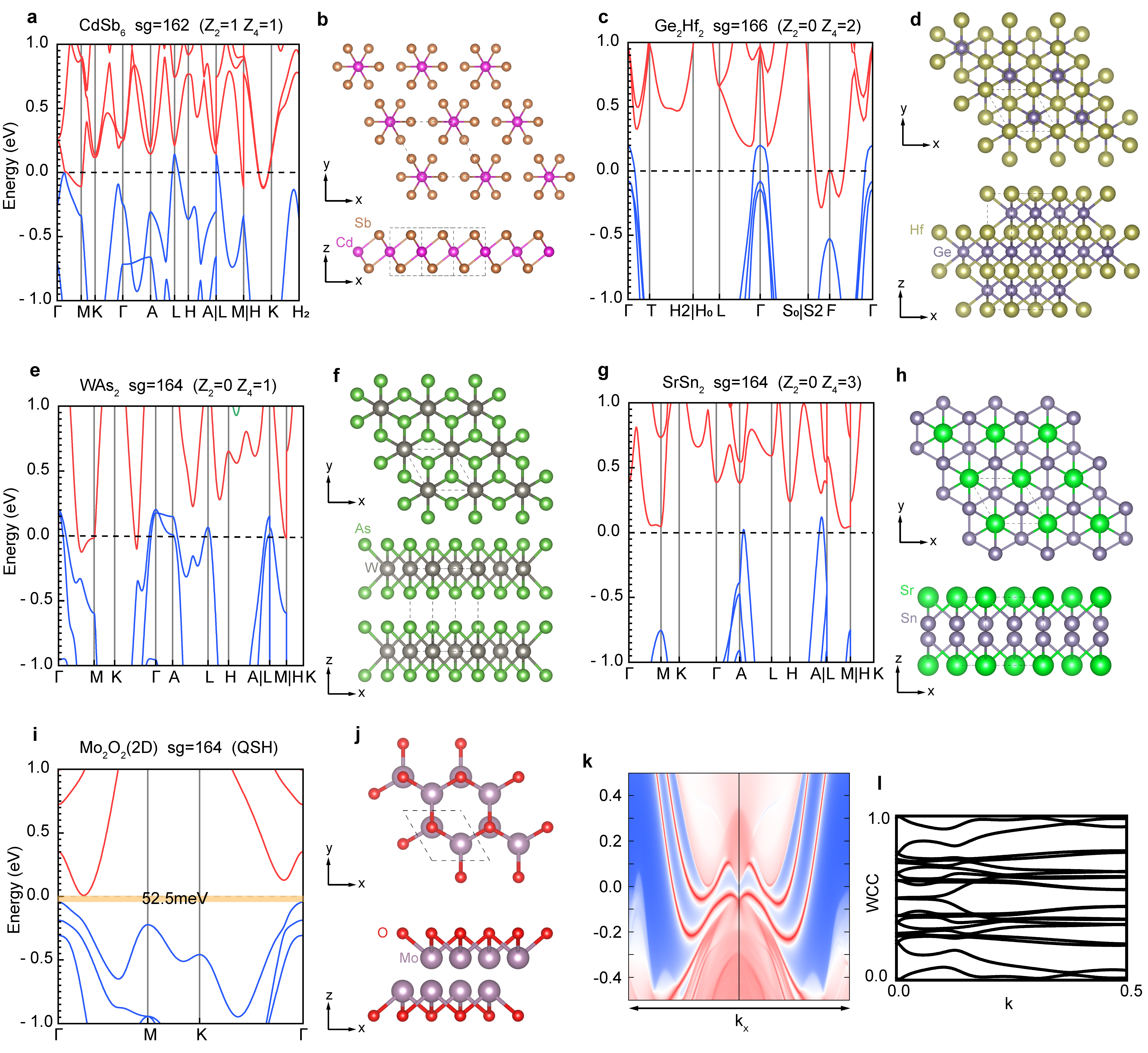}
\end{center}
\caption{\textbf{Five representative TIs and TCIs exhibiting simple and clean band struture near the Fermi level.} \textbf{a}–\textbf{j}, Band structures and crystal structures of the selected materials including CdSb$_{6}$(\textbf{a}-\textbf{b}), Ge$_{2}$Hf$_{2}$(\textbf{c}-\textbf{d}), WAs$_{}$(\textbf{e}-\textbf{f}), SrSn$_{2}$(\textbf{g}-\textbf{h}), and Mo$_{2}$O$_{2}$(\textbf{i}-\textbf{j}). \textbf{k}-\textbf{l}, Edge states and Wannier charge centers of Mo$_{2}$O$_{2}$.}
\label{fig4}
\end{figure*}

\begin{table*}[htbp]
	\caption{\textbf{TIs and TCIs with clean and simple structures near the Fermi level, generated by the DC+XB model.} The table lists their chemical formulas, space groups, and topological index, which are computed using the framework described in Ref.~\cite{gaojiacheng2022}.}
    \begin{center}
		\renewcommand{\arraystretch}{1.2}
		\begin{tabular*}{3.3in}
			{@{\extracolsep{\fill}}c|cc}
			\hline
			\hline
			Materials & Space Group & Topological Index\\
			\hline
            CdSb$_{6}$ & 162 & Z$_{2}$=1 Z$_{4}$=1 \\
            Ge$_{2}$Hf$_{2}$ & 166 & Z$_{2}$=0 Z$_{4}$=2 \\
            WAs$_{2}$ & 164 & Z$_{2}$=0 Z$_{4}$=1 \\
            SrSn$_{2}$ & 164 & Z$_{2}$=0 Z$_{4}$=3 \\
            Mo$_{2}$O$_{2}$ (2D) & 164 & QSH \\
            Hf$_{7}$Sn$_{2}$ & 162 & Z$_{2}$=1 Z$_{4}$=0 \\
            PbSn$_{2}$K$_{6}$ & 162 &Z$_{2}$=1 Z$_{4}$=3 \\
            La$_{2}$C$_{6}$Si$_{2}$Li$_{6}$ & 162 & Z$_{2}$=0 Z$_{4}$=1 \\
			MoP$_{2}$ & 164 & Z$_{2}$=1 Z$_{4}$=0 \\
            GeY$_{2}$Au$_{2}$ & 164 & Z$_{2}$=0 Z$_{4}$=2 \\
            Zr$_{6}$Pb$_{6}$Si$_{6}$ & 166 & Z$_{2}$=1 Z$_{4}$=0 \\
            O$_{3}$Pb$_{3}$Hf$_{6}$ & 166 & Z$_{2}$=1 Z$_{4}$=1 \\
            Hf$_{3}$Ni$_{3}$In$_{6}$ & 166 & Z$_{2}$=1 Z$_{4}$=3 \\
            Ni$_{9}$Pb$_{3}$Ge$_{6}$Y$_{6}$ & 166 & Z$_{2}$=0 Z$_{4}$=3 \\
			\hline
			\hline
		\end{tabular*}
	\end{center}
\end{table*}

\subsection*{Material generation}
To further demonstrate the effectiveness of the DC+XB model, we generate a total of 4,480 materials and search for novel TI and TCI candidates validated through first-principles calculations. In particular, we focus on materials belonging to space groups 162, 164 and 166 for the following reasons. First, based on prior knowledge, many promising TIs are typically found in the trigonal crystal system, motivating our decision to constrain generation within this class~\cite{Zhang2009, zhangjianmin2013}. Furthermore, as reported in Ref.~\cite{xu2024c}, these space groups are statistically more likely to host topological phases. They also possess high crystallographic symmetry and are associated with non-trivial symmetry indicator groups, enabling unambiguous identification of topological properties via symmetry indicators.

After generation, we first removed all materials already present in the Materials Project database and retained only those predicted by XBERT to be TIs and TCIs. We applied additional filtering based on materials science knowledge to identify materials that are more feasible for synthesis in experimental settings. Specifically, we excluded materials containing rare or unstable elements, such as those with atomic numbers greater than 84 or certain lanthanides. We also constrained the number of distinct elements to two, three, or four, thereby favoring chemically simpler compounds. Furthermore, to increase the likelihood of obtaining materials with insulating bulk states, we required the presence of at least one $p$-block element. 
As a result, a total of 493 candidate materials were subjected to structural relaxation and band structure calculations without spin-orbit coupling (SOC), with approximately $90\%$ of them successfully converging during the relaxation process, suggesting that most generated materials are indeed stable.

Although it is in principle straightforward to perform detailed topological characterization and dynamical stability analysis on all such candidates, this approach is computationally expensive and inefficient. To address this, we further refined our candidate set to identify high-quality TIs and TCIs that are both experimentally relevant and computationally tractable. Specifically, we focused on nonmagnetic materials exhibiting clean and simple band structures near the Fermi level after structural relaxation. Through this process, we identified a total of 15 new TI and TCI materials. Among them, a particularly notable discovery is Ge$_2$Bi$_2$O$_6$, which we identify as a strong TI featuring a topologically non-trivial full band gap of 255 meV. This gap ranks among the largest of all materials in the topological materials database~\cite{Vergniory2019}, as shown in Fig.~1e. Ge$_2$Bi$_2$O$_6$ has a hexagonal lattice with space group $P\bar{3}1m$ (No. 162), as shown in Fig.~1c, and its bulk band structure and edge state spectrum are shown in Fig.~1f and Fig.~1g, respectively. Furthermore, we computed its phonon spectrum and Wilson loop, provided in the Supplementary Information, which confirm both its dynamical stability and topological non-triviality. It is worth emphasizing that most entries in existing topological materials databases correspond to metallic compounds with non-trivial symmetry indicators, rather than insulators with a clean bulk gap. Therefore, the discovery of a strong TI with a large band gap such as Ge$_2$Bi$_2$O$_6$ is both significant and promising for practical applications. In addition, through elemental substitution, we obtained a related compound, Sn$_2$Bi$_2$O$_6$, which also exhibits a non-trivial full band gap of 116.6 meV, further suggesting that these materials may constitute a new family of strong TIs, comparable in performance to the most famous Bi$_2$Se$_3$-type materials. The remaining 14 TI and TCI materials identified are listed in Table 2, with their corresponding band structures and crystal structures shown in Fig.~4 and the Supplementary Information. Among the materials shown in Fig.~4, we highlight Mo$_{2}$O$_{2}$ as a two-dimensional quantum spin Hall (QSH) insulator with a full band gap of 52.5 meV. Notably, although the generative model is designed to directly generate three-dimensional structures, Mo$_{2}$O$_{2}$ exhibits clear two-dimensional characteristics, as it consists of stacked layers with weak interlayer coupling. By extending the lattice constant along the $z$-direction, we effectively treat the material as a 2D system. The resulting nontrivial edge states and Wannier charge centers are shown in Fig.~4k and 4l. These findings collectively demonstrate the effectiveness of the ReFT framework in generating high-quality materials with targeted physical properties.

\section*{Discussion}
This work demonstrates that topological materials, particularly TIs and TCIs, can be effectively generated using the  ReFT technique applied to a generative model. To the best of our knowledge, this is the first study to integrate the ReFT technique into the material generation process, and also the first to focus specifically on the generation of topological materials—properties that are more intricate and complex than those explored in previous studies, such as formation energy and band gap~\cite{Zeni2025}. Given the unlimited potential for material generation using this fine-tuned model, we expect a wide variety of interesting materials to be produced, with the primary constraint being the computational cost of first-principles calculations to verify the topological properties.

One limitation of the present work is that many of the generated TIs and TCIs exhibit vanishing band gaps, whereas the most practically valuable materials are those with a sizable full band gap. A promising direction for future research is to explore the incorporation of predicted band gap information into the reward function, thereby guiding the model toward generating topological materials with more desirable insulating properties. Another potential avenue is the integration of SFT and ReFT. Given that SFT has demonstrated effectiveness in generating materials with targeted band gaps~\cite{Zeni2025}, applying ReFT subsequently could further refine the generation process to yield fully gapped topological materials. Moreover, this combined strategy may also offer a promising pathway for the discovery of magnetic TIs, a class of materials that remains challenging to realize~\cite{Zhangd2019, YujunDeng2020, xu2020, bernevig2022}. Finally, we note that the ReFT framework is not limited to topological materials. Its generality makes it a compelling approach for the design of materials with other complex or rare properties, such as high superconducting transition temperatures. We leave these directions for future investigation.

\section*{Methods}
\subsection*{Form of $p_{\theta}(\mathcal{M}_{t-1}|\mathcal{M}_{t})$}

To illustrate the material generation process, we consider a material consisting of $N$ atoms distributed across $N^{\prime}$ distinct Wyckoff positions.  We begin by analyzing the generation of atomic species $\textbf{\textit{A}}\in\mathbb{R}^{h\times N}$. Given that all atoms within the same set of Wyckoff positions share an identical atomic type, it suffices to generate only the basic atomic species $\textbf{\textit{A}}^{\prime} \subseteq \textbf{\textit{A}}$, where $\textbf{\textit{A}}^{\prime} \in \mathbb{R}^{h\times N^{\prime}}$ represents a continuous one-hot encoding of atomic types. The generation of  $\textbf{\textit{A}}^{\prime}$ follows the standard Denoising Diffusion Probabilistic Model (DDPM) with the forward noise addition process formulated as $q(\textbf{\textit{A}}_{t}^{\prime}|\textbf{\textit{A}}_{0}^{\prime})=\mathcal{N}\left(  \textbf{\textit{A}}_{t}^{\prime}|\sqrt{\bar{\alpha}_{t}}\textbf{\textit{A}}_{0}^{\prime},(1-\bar{\alpha}_{t})\textbf{\textit{I}} \right)$, which in turn defines the reverse generation process: $p_{\theta}(\textbf{\textit{A}}_{t-1}^{\prime}|\mathcal{M}_{t})=\mathcal{N}\left(  \textbf{\textit{A}}_{t-1}^{\prime}|\mu_{\textbf{\textit{A}}^{\prime}}(\mathcal{M}_{t}),\beta_{t}\frac{1-\bar{\alpha}_{t-1}}{1-\bar{\alpha}_{t}}\textbf{\textit{I}}\right)$, where $\mathcal{N}(\cdot,\cdot )$ represents a normal distribution, $\mu_{\textbf{\textit{A}}^{\prime}}(\mathcal{M}_{t})=\frac{1}{\sqrt{\alpha_{t}}}\left( \textbf{\textit{A}}_{t}^{\prime}-\frac{\beta_{t}}{\sqrt{1-\bar{\alpha}_{t}}}\hat{\mathbf{\mathit{\epsilon}}}_{\textbf{\textit{A}}^{\prime}}(\mathcal{M}_{t},t)
 \right)$ and the term $\hat{\mathbf{\mathit{\epsilon}}}_{\textbf{\textit{A}}^{\prime}}(\mathcal{M}_{t},t) \in \mathbb{R}^{h\times N^{\prime}}$ is predicted by the denoising model $\phi_{\theta}(\mathcal{M}_{t},t)$. Here, $\bar{\alpha}_{t}$ and $\beta_{t}$ are the variance of each diffusion step controlled by the cosine scheduler adopted in Ref.~\cite{jiao2024}. The generation of the lattice matrix $\textbf{\textit{L}}\in\mathbb{R}^{3\times 3}$ closely parallels that of $\textbf{\textit{A}}$, with a key distinction: to facilitate the enforcement of space group constraints on the lattice structure, we generate $\textbf{\textit{k}}=(k_{1},k_{2},...,k_{6})$ instead of directly generating $\textbf{\textit{L}}$ as discussed in the main text. The forward process follows $q(\textbf{\textit{k}}_{t}|\textbf{\textit{k}}_{0})=\mathcal{N}\left(  \textbf{\textit{k}}_{t}|\sqrt{\bar{\alpha}_{t}}\textbf{\textit{k}}_{0},(1-\bar{\alpha}_{t})\textbf{\textit{I}} \right)$, while the reverse process is given by $p_{\theta}(\textbf{\textit{k}}_{t-1}|\mathcal{M}_{t})=\mathcal{N}\left(  \textbf{\textit{k}}_{t-1}|\mu_{\textbf{\textit{k}}}(\mathcal{M}_{t}),\beta_{t}\frac{1-\bar{\alpha}_{t-1}}{1-\bar{\alpha}_{t}}\textbf{\textit{I}}\right)$. Similarly, $\mu_{\textbf{\textit{k}}}(\mathcal{M}_{t})=\frac{1}{\sqrt{\alpha_{t}}}\left( \textbf{\textit{k}}_{t}-\frac{\beta_{t}}{\sqrt{1-\bar{\alpha}_{t}}}\hat{\mathbf{\mathit{\epsilon}}}_{\textbf{\textit{k}}}(\mathcal{M}_{t},t)
 \right)$ and the term $\hat{\mathbf{\mathit{\epsilon}}}_{\textbf{\textit{k}}}(\mathcal{M}_{t},t) $ is predicted by the model $\phi_{\theta}(\mathcal{M}_{t},t)$. 
At each step of generation, $\textbf{\textit{k}}$ is explicitly adjusted to conform to the space group requirements. The generation of fractional coordinates $\textbf{\textit{F}}\in\mathbb{R}^{3\times N}$ presents a greater challenge, as it must account for their inherent periodic translation invariance, a property that necessitates the adoption of the Score-Matching framework. Similar to the treatment of $\textbf{\textit{A}}$, it is only necessary to determine the fractional coordinates of the basic atom within a Wyckoff position, since the coordinates of all other atoms within the same position can be uniquely determined based on symmetry constraints. Consequently, rather than modeling the full set of fractional coordinates, our focus is exclusively on  $\textbf{\textit{F}}^{\prime}\in\mathbb{R}^{3\times N^{\prime}}$, which serves as the fundamental representation from which the complete atomic arrangement can be inferred. The forward process is conducted via the wrapped normal distribution: $q(\textbf{\textit{F}}_{t}^{\prime}|\textbf{\textit{F}}_{0}^{\prime})=\mathcal{N}_{w}\left(  \textbf{\textit{F}}_{t}^{\prime}|\textbf{\textit{F}}_{0}^{\prime},\sigma^{2}_{t}\textbf{\textit{I}} \right)$ and the backward process is implemented using the denoising term $\hat{\mathbf{\mathit{\epsilon}}}_{\textbf{\textit{F}}^{\prime}}(\mathcal{M}_{t},t)$ produced by the model $\phi_{\theta}(\mathcal{M}_{t},t)$. The exact form of  $p_{\theta}(\textbf{\textit{F}}_{t-1}^{\prime}|\mathcal{M}_{t})$  is relatively complex and is detailed as follows. 

To simplify the problem, we consider the update step for a single basic atom, transitioning from $\textbf{\textit{x}}_t$ to $\textbf{\textit{x}}_{t-1}$, where $\textbf{\textit{x}}_{t}, \textbf{\textit{x}}_{t-1}\in \mathbb{R}^{3}$. The generalization to multiple basic atoms is straightforward. The backward sampling process is given by $\textbf{\textit{x}}_{t-1} = (\textbf{\textit{x}}_{t}+\hat{\boldsymbol{\epsilon}}_{t}) + \xi_{t}\boldsymbol{\epsilon}$, where $\hat{\boldsymbol{\epsilon}}_{t}\in \mathbb{R}^{3}$ is predicted by the denoising model, $\xi_{t}$ is a hyperparameter determined by the noise specified during the forward noising process as in Ref.~\cite{jiao2024}, and $\boldsymbol{\epsilon}\in \mathbb{R}^{3}$ is drawn from the standard normal distribution. For simplicity of notation, certain coefficients have been omitted. The corresponding probability is $p_{\theta}(\textbf{\textit{x}}_{t-1}|\textbf{\textit{x}}_{t}) = \frac{1}{\sqrt{(2\pi\xi_{t}^{2})^{3}}}\text{exp}\left(-\frac{\|\textbf{\textit{x}}_{t-1}-\textbf{\textit{x}}_{t}-\hat{\boldsymbol{\epsilon}}_{t}\|^2}{2\xi_{t}^{2}}\right)$, which is a standard normal distribution and $||\cdot||$ denotes the vector norm. Further, if the sampling process incorporates a truncation function, such that it transforms into $\textbf{\textit{x}}_{t-1} = w\left((\textbf{\textit{x}}_{t}+\hat{\boldsymbol{\epsilon}}_{t}) + \xi_{t}\boldsymbol{\epsilon}\right)$, where $w(\cdot)$ retains the fractional part of the input, then the corresponding probability distribution becomes $p_{\theta}(\textbf{\textit{x}}_{t-1}|\textbf{\textit{x}}_{t}) \propto \sum_{\textbf{\textit{z}}\in \mathbb{Z}^{3}}\text{exp}\left(-\frac{\|\textbf{\textit{x}}_{t-1}-\textbf{\textit{x}}_{t}-\hat{\boldsymbol{\epsilon}}_{t} + \textbf{\textit{z}}\|^2}{2\xi_{t}^{2}}\right)$, which is referred to as the wrapped normal distribution. In DiffCSP++, to ensure that the generated atomic coordinates adhere to the symmetry constraints of Wyckoff positions, $\textbf{\textit{F}}_{t}^{\prime}$ is updated in the following more complicated form: $\textbf{\textit{x}}_{t-1} = w\left(A\left((\textbf{\textit{x}}_{t}+\hat{\boldsymbol{\epsilon}}_{t}) + \xi_{t}\boldsymbol{\epsilon}\right)+\textbf{\textit{b}}\right)$, where $A$ is a $3 \times 3$ matrix and $\textbf{\textit{b}}$ is a  $3 \times 1$ vector that projects and shifts the updated location in accordance with the specific Wyckoff position. Thus, after being updated according to the standard normal distribution, a linear transformation is applied, followed by the function $w(\cdot)$ to produce the final output. We can further rewrite the sampling process as $\textbf{\textit{x}}_{t-1} = w\left(\textbf{\textit{m}} + \xi_{t}A\boldsymbol{\epsilon}\right)$ with $\textbf{\textit{m}} \equiv A(\textbf{\textit{x}}_{t}+\hat{\boldsymbol{\epsilon}}_{t})+\textbf{\textit{b}}$ is a $3$-dimensional vector, and 
\begin{eqnarray}
\xi_{t}A\boldsymbol{\epsilon}&=&\textstyle\begin{bmatrix} n_{11} \ \ \ n_{12} \ \ \ n_{13} \\n_{21} \ \ \ n_{22} \ \ \ n_{23} \\n_{31} \ \ \ n_{32} \ \ \ n_{33} \end{bmatrix}\textstyle\begin{bmatrix} \epsilon_{1} \\ \epsilon_{2} \\ \epsilon_{3} \end{bmatrix} \\ \nonumber
&=&\textstyle\begin{bmatrix} \sqrt{n_{11}^{2}+n_{12}^{2}+n_{13}^{2}}\epsilon_{1}^{\prime} \\ \sqrt{n_{21}^{2}+n_{22}^{2}+n_{23}^{2}}\epsilon_{2}^{\prime}  \\ \sqrt{n_{31}^{2}+n_{32}^{2}+n_{33}^{2}}\epsilon_{3}^{\prime}  \end{bmatrix},
\end{eqnarray}
where $\epsilon_{1},\epsilon_{2},\epsilon_{3},\epsilon_{1}^{\prime},\epsilon_{2}^{\prime},\epsilon_{3}^{\prime}$ are all sampled from the standard normal distribution. Thus, we have $x_{t-1,i}=w(m_{i}+\sqrt{n_{i1}^{2}+n_{i2}^{2}+n_{i3}^{2}}\epsilon_{i}^{\prime})$, corresponding to the probability $p_{\theta}(x_{t-1,i}|\mathcal{M}_{t})\propto \sum_{\textbf{\textit{z}}\in\mathbb{Z}^3}\text{exp}\left(-\frac{(x_{t-1}-m_{i} + z)^2}{2(n_{i1}^2+n_{i2}^2+n_{i3}^2)}\right)$, with $p_{\theta}(\textbf{\textit{x}}_{t-1}|\mathcal{M}_t)=\prod_{i=1}^{3}p_{\theta}(x_{t-1,i}|\mathcal{M}_t)$. It is also worth mentioning that for positions in the Wyckoff positions with fixed coordinates, we assign them a probability of one during update. In the actual implementation, the transition from $\textbf{\textit{F}}_{t}^{\prime}$ to $\textbf{\textit{F}}_{t-1}^{\prime}$ is performed using a predictor-corrector sampler, i.e., first evolving $\textbf{\textit{F}}_{t}^{\prime}$ to $\textbf{\textit{F}}_{t-\frac{1}{2}}^{\prime}$, and then from $\textbf{\textit{F}}_{t-\frac{1}{2}}^{\prime}$ to $\textbf{\textit{F}}_{t-1}^{\prime}$. Each step follows the same formulation as the previously described transition from $\textbf{\textit{x}}_{t}$ to $\textbf{\textit{x}}_{t-1}$, so the total transition probability can be obtained by computing the probabilities of the two steps and multiplying them together.

\subsection*{Objective function}
Using the PPO algorithm, our objective function is ultimately formulated as:
\begin{align}
    J^{\text{PPO}}(\theta) = \mathbb{E}_{\tau \sim p_{\theta'}}\Bigg[ \sum_{t=1}^{T} \text{min}\Bigg( 
    \frac{p_{\theta}(\mathcal{M}_{t-1}|\mathcal{M}_{t})}{p_{\theta^{\prime}}(\mathcal{M}_{t-1}|\mathcal{M}_{t})}r(\mathcal{M}_{0}),\notag \\ \quad \text{clip}\left( \frac{p_{\theta}(\mathcal{M}_{t-1}|\mathcal{M}_{t})}{p_{\theta^{\prime}}(\mathcal{M}_{t-1}|\mathcal{M}_{t})},1-\epsilon,1+\epsilon \right)r(\mathcal{M}_{0})
    \Bigg) \Bigg],
\end{align}
where the expectation is taken over denoising trajectories generated by the parameters $\theta^{\prime}$, $\epsilon$ is a small hyperparameter, and the $\text{clip}$ function restricts the probability ratio within the interval $[1-\epsilon, 1+\epsilon]$. In this work, we set $\epsilon$ to 10$^{-4}$.

\subsection*{Hyperparameters and training details}
We trained our model using PyTorch on an NVIDIA 6000 Ada GPU. In each batch, we randomly generate 128 materials, with each material undergoing a diffusion process of 1000 steps. A total of 30 batches were generated for reinforcement learning. The learning rate was set to 0.00014. We found that increasing the learning rate or the number of batches excessively could further improve the proportion of TIs. However, it would also lead to a more significant decline in other evaluation metrics for material generation. Therefore, we adopted relatively conservative hyperparameters to strike a balance.

\subsection*{DFT details}
The first-principles calculations are carried out in the framework of the generalized gradient approximation (GGA) functional of the density functional theory through employing the Vienna \emph{ab initio} simulation package (VASP) with projector augmented wave pseudopotentials~\cite{Kresse1996,Bl1994,Perdew1996}. The lattice constants and inner positions are obtained through full relaxation with a force tolerance criterion for convergence of 0.01 eV/\AA. The convergence criterion for the total energy was 10$^{-6}$ eV. The SOC effect is self-consistently included. By considering the transition metal, LDA+U functional with different U values are adopted~\cite{Dudarev1998}. Surface state calculations, namely the LDOS and Wannier charge center are performed based on maximally localized Wannier functions by Wannier90~\cite{Arash2008} and the WannierTools packages~\cite{QuanSheng2018}. The open-source program Irvsp~\cite{Jiacheng2021} is adopted for irreducible representations analysis and the workflow discribed in Ref.~\cite{gaojiacheng2022} is used to determine the topological nature.

\section*{Acknowledgment}
This work is supported by the Natural Science Foundation of China through Grants No.~12350404 and No.~12174066, the Innovation Program for Quantum Science and Technology through Grant No.~2021ZD0302600, the Science and Technology Commission of Shanghai Municipality under Grants No.~23JC1400600, No.~24LZ1400100, and No.~2019SHZDZX01. Y.J. acknowledges the support from the Postdoctoral Fellowship Program of CPSF under No.~GZC20240302 and No.~2024M760488. 

J.W. supervised the project. H.X. trained both the XBERT and DC+XB models, evaluated their performance, and generated materials. D.Q. proposed and developed the use of ReFT in generating topological materials. H.X., Z.L., and Y.J. performed the DFT calculations. All authors contributed to the analysis and discussion of the results. H.X., D.Q. and J.W. wrote the paper with the contribution of all authors. The authors declare no competing interests. All raw data necessary for reproducing the figures in the manuscript, including the crystallographic information files of the discovered topological materials, and the DC+XB and XBERT models will be made accessible in a GitHub repository upon publication. Additional explanations and details regarding the model and generated materials can be found in the Supplementary Information.

%

\end{document}


\title{Supplementary Information for ``Design Topological Materials by Reinforcement Fine-Tuned Generative Model"}


\author[1,2]{\fnm{Haosheng} \sur{Xu}}
\equalcont{These authors contributed equally to this work.}

\author[1,2]{\fnm{Dongheng} \sur{Qian}}
\equalcont{These authors contributed equally to this work.}

\author[1,2]{\fnm{Zhixuan} \sur{Liu}}

\author[1,2]{\fnm{Yadong} \sur{Jiang}}

\author*[1,2,3,4]{\fnm{Jing} \sur{Wang}}\email{wjingphys@fudan.edu.cn}

\affil[1]{\orgdiv{State Key Laboratory of Surface Physics and Department of Physics}, \orgname{Fudan University}, \orgaddress{\city{Shanghai}, \postcode{200433}, \country{China}}}

\affil[2]{\orgname{Shanghai Research Center for Quantum Sciences}, \orgaddress{\city{Shanghai}, \postcode{201315}, \country{China}}}

\affil[3]{\orgdiv{Institute for Nanoelectronic Devices and Quantum Computing}, \orgname{Fudan University}, \orgaddress{\city{Shanghai}, \postcode{200433},  \country{China}}}

\affil[4]{\orgname{Hefei National Laboratory}, \orgaddress{\city{Hefei}, \postcode{230088}, \country{China}}}

\maketitle

\textbf{This PDF file includes:}

Supplementary Text

Table. S1

Figs. S1 to S3


\newpage

\section{Performance of the DC+XB model under alternative parameter settings}

We have conducted a series of experiments to assess the impact of hyperparameter selection. One of the representative configurations tested involved increasing the learning rate to 0.0003, setting the batch size to 256, and training the model for 20 batches. Within each batch, the reward for each material was normalized by subtracting the mean reward of that batch. The resulting model, referred to as DC+XB(AP), is characterized by its topological material proportions and evaluation metrics, as summarized in Table S1. It is observed that the proportions of topological insulators (TIs) and topological crystalline insulators (TCIs) are higher than in DC+XB, while the proportion of trivial materials is lower. However, a more pronounced decline in stability metrics is noted. Given that stability is rigorously verified through the calculation of the phonon spectrum—a computation more expensive than those evaluating topological properties—we have opted to sacrifice some of the increase in topological materials proportion in favor of achieving higher stability. Additionally, it is observed that setting the number of batches too high may lead to unreliable results, potentially causing errors during the material generation process.

\section{More information on the generated topological materials} 

For the highlighted compound Ge$_{2}$Bi$_{2}$O$_{6}$ in the main text, we further calculated its Wannier charge centers (WCCs) and phonon spectrum. The phonon calculations were carried out using the finite displacement method, as implemented in the Phonopy package~\cite{TOGO20151}, with a supercell size of 4$\times$4$\times$2. Spin-orbit coupling was included in all calculations. As shown in Fig.~S1, the evolution of the WCCs reveals a Fu-Kane topological invariant of (1;000). Furthermore, the absence of imaginary frequencies in the phonon spectrum confirms the material's dynamical stability. Fig.~S2 presents the phonon spectra for the materials listed in Fig.~4 of the main text. These spectra were also calculated using the finite displacement method within the Phonopy package. The absence of imaginary frequencies indicates that all these materials are dynamically stable. Fig.~S3 displays the calculated electronic band structures, corresponding space groups, and topological indices for the other materials listed in Table 2 of the main text, validating their non-trivial topological properties.



\clearpage

\newpage

\begin{table*}[htbp]
	\caption{\textbf{The proportions of different types of materials and the metrics for DC+XB(AP).} The stability metrics were evaluated on a set of 1,280 generated materials, rather than on 10,000 samples as in previous evaluations.} 
    \begin{center}
		\renewcommand{\arraystretch}{2}
		\begin{tabular*}{3in}
			{@{\extracolsep{\fill}}c|cc}
			\hline
			\hline
			percentage of Trivial & 21.56\% & \\
            percentage of TI(TCI) & 33.20\% & \\
            percentage of TSM & 45.23\% & \\
            \hline
            struc. & 100.00 & \\
            comp. & 82.11 & \\
            cov-R  & 90.64 & \\
            cov-P & 93.28 & \\
            $d_{\rho}$ & 2.30 & \\
            $d_{elem}$ & 0.6448 & \\
            
			\hline
			\hline
		\end{tabular*}
	\end{center}
\end{table*}

\begin{figure*}[htbp]
\begin{center}
\includegraphics[width=5in, clip=true]{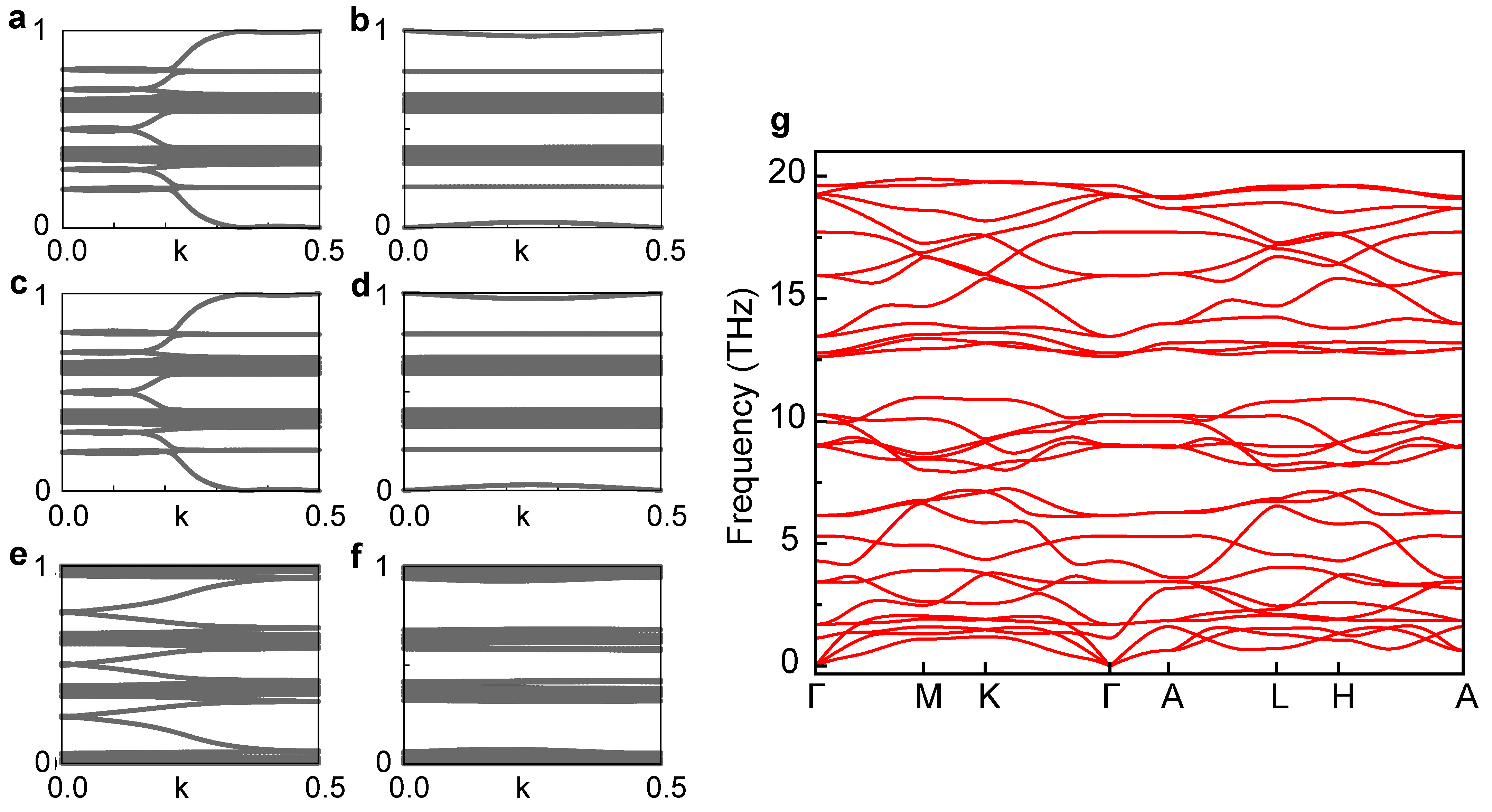}
\end{center}
\caption{Wannier charge center (WCC) and phonon spectrum of Ge$_{2}$Sn$_{2}$O$_{6}$.}
\label{figS1}
\end{figure*}

\begin{figure*}[htbp]
\begin{center}
\includegraphics[width=5in, clip=true]{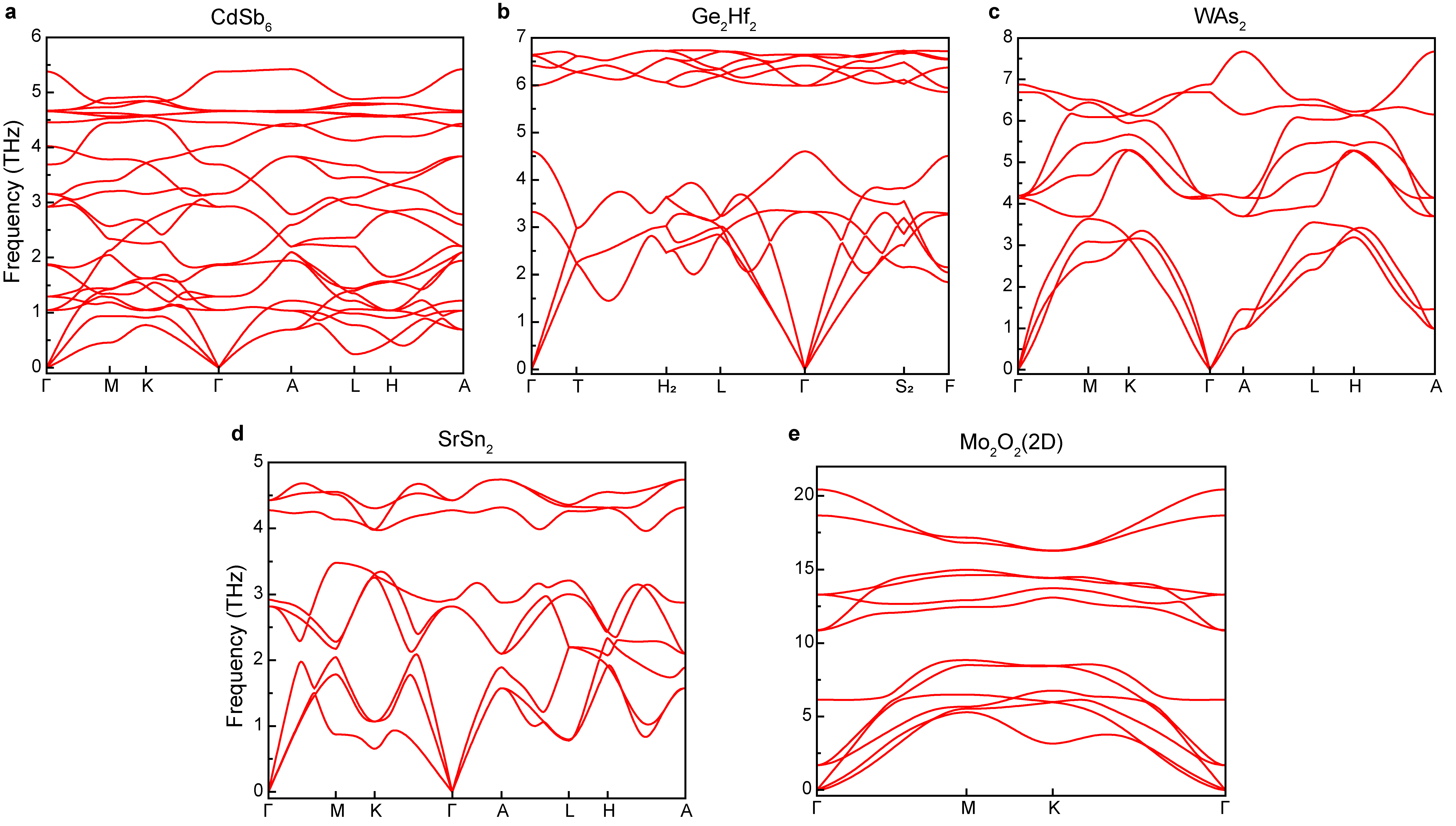}
\end{center}
\caption{Phonon spectra of the materials in Fig.~4 of the main text.}
\label{figS2}
\end{figure*}

\begin{figure*}[htbp]
\begin{center}
\includegraphics[width=5in, clip=true]{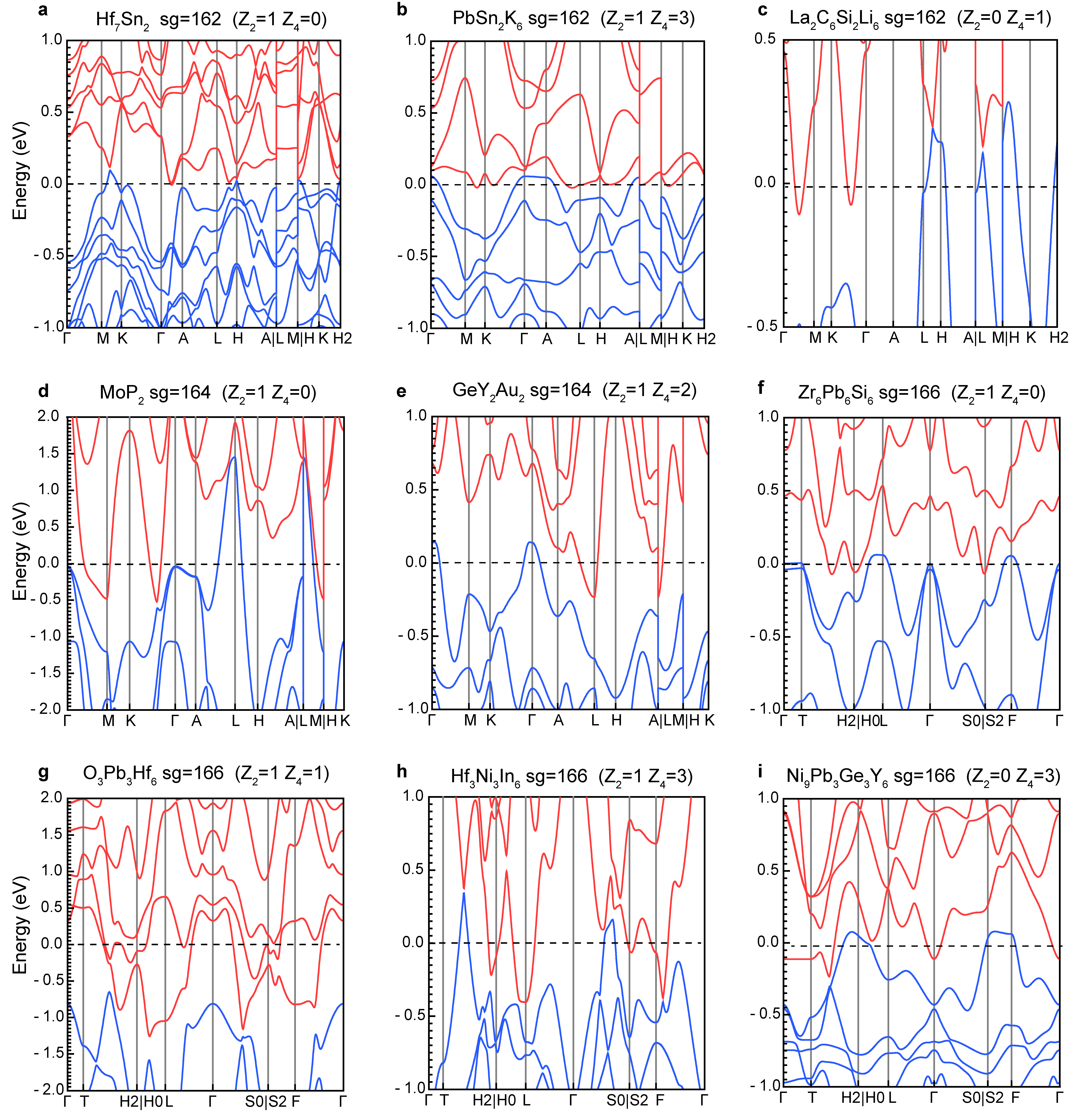}
\end{center}
\caption{Band structures, space group, and topological index of other materials in Table 2 of the main text.}
\label{figS3}
\end{figure*}